\documentclass{appolb}
\usepackage{epsf,rotate}
\usepackage{amsmath,amstext,amsbsy,amsopn,amsfonts,amssymb}
\usepackage[dvips]{graphicx}
\usepackage[latin1]{inputenc}
\usepackage[english]{babel}
\usepackage[T1]{fontenc}
\usepackage{psfrag}

\newcommand{\eval}[2][\right]{\relax
  \ifx#1\right\relax \left.\fi#2#1\rvert}

\newcommand{\as}{\ifmmode\alpha_{\rm s}\else{$\alpha_{\rm s}$}\fi}
\newcommand{\asbar}{\ifmmode\wbar {\alpha}_{\rm s}\else{$\wbar{\alpha}_{\rm s}$}\fi}

 \catcode`\@=11

\def \lab #1 {\label{#1}}
\newcommand \ci [1] {\cite{#1}}

\def\r{{r}}
\def\c{{\hat{c}}}

\def\hph0{{\hphantom{0}}}

\begin{document}

\eqsec
\bibliographystyle{unsrt}

\title{Virial theorem for four-dimensional 
supersymmetric Yang-Mills quantum mechanics with $SU(2)$ 
gauge group\footnote{Presented at the Cracow School of Theoretical Physics,
Zakopane 2006}}

\author{{\bf Jan Kota{\'n}ski}
\address{M. Smoluchowski Institute of Physics, Jagellonian University\\
Reymonta 4, 30-059 Krak{\'o}w, Poland}}

\maketitle

\begin{abstract}
{\normalsize 
Supersymmetric Yang-Mills quantum mechanics (SYMQM) 
in four dimensions
for $SU(2)$ gauge group is considered. 
In this work a two-fermionic sector with the angular momentum 
$j=0$ in discussed. Energy levels from discrete 
and continuous spectra are calculated.
To distinguish localized states from non-localized ones
the virial theorem is applied.}
\end{abstract}


\newline   

\noindent TPJU-12/2006 \newline


\section{Introduction}

In this work we consider supersymmetric Yang-Mills quantum mechanics (SYMQM)
\ci{Witten:1981nf,Claudson:1984th}. 
Theory for $SU(N \to \infty)$ gauge group in $D=10$
dimensions is related to $M-$theory
and allows researches on the BFSS hypothesis
\ci{Banks:1996vh}.
For a smaller number of dimensions and a smaller number of colours,
like in our case where $D=4$ and $N=2$ respectively,
such mechanics pose excellent theoretical laboratory 
\ci{Wosiek:2002nm}
to test amazing properties of supersymmetry, \ie
coexistence of discrete and continuous spectrum 
\ci{deWit:1988ct},
action of supersymmetric generators or unique  features
of SUSY vacuum states 
\ci{Claudson:1984th,Halpern:1997fv,Polchinski:1998rq,vanBaal:2001ac}. 
Moreover SYMQM without fermions
describe glueballs which are also considered in many 
non-supersymmetric theories 
\ci{Luscher:1982ma,
vanBaal:1988qm,Wosiek:2002nm}.
Here, in order to calculate the energy 
spectrum of our model the method proposed by
van Baal in Ref. \ci{vanBaal:2001ac} is used.
In this work eigenvalues as well as eigenfunctions of Hamiltonian
of the model are computed. Finally, to distinguish the localized states form 
the non-localized ones the virial theorem is applied.

\section{Hamiltonian}

The zero-volume Hamiltonian \ci{Luscher:1982ma,vanBaal:2001ac},
\ie in the long wave approximation \ci{Claudson:1984th,Halpern:1997fv},
reads 
\begin{equation}
 H =-
\underbrace{\underbrace{
\frac{1}{2}
\sum_{i,a} \left( \frac{\partial }{\partial 
\hat{c}_{i}^{a}}\right) ^{2}
\rule[-0.5cm]{0cm}{1cm}
}_{H_T}
+
\underbrace{\frac{1}{2}
\sum_{i,a} \left( \hat{B}_{i}^{a}\right) ^{2}
\rule[-0.5cm]{0cm}{1cm}
}_{H_V}
\rule[-1.cm]{0cm}{1cm}
}_{H_B}-
\underbrace{i
\sum_{i,a,b,d} 
\varepsilon _{abd}\bar{\lambda }^{a}\bar{\sigma }^{i}
\lambda ^{b}\hat{c}_{i}^{d}\,,
\rule[-0.5cm]{0cm}{1cm}
}_{H_F}
\lab{eq:Ham}
\end{equation}  
where in the bosonic part, \ie in $H_B$,  
$H_T$ is the kinetic Hamiltonian, $H_V$ with 
\begin{equation}
\hat{B}_{i}^{a}
=-\frac{1}{2}
\sum_{i,j,k,a,b,d} 
\varepsilon _{ijk}
\varepsilon _{abd} \c_{j}^{b} \c_{k}^{d}\,,
\lab{eq:BB}
\end{equation}
corresponds to 
a bosonic potential
while the fermionic part is denoted by $H_F$, where
$\sigma ^{j}=\tau^j$ are Pauli matrices.
Bosonic variables, $\c_i^a$, have colour indices $a=1,2,3$
and spatial indices $i=1,2,3$. 
Except the colour indices the anticommuting Weyl spinors, $\lambda_a^{\alpha}$,
have spinor indices $\alpha=1,2$.
Thus, in this system the maximum number of fermions is $3 \times 2=6$.

\section{The cut Fock space}

The operators of the number of fermionic quanta, $n_F$,
and the total angular momentum of the system, $j$ commute with the Hamiltonian.
Therefore, solving the eigenproblem of the Hamiltonian we can 
separately consider the sectors with fixed values of $n_F$ and $j$.
The system has particle-hole symmetry so it is enough to consider sectors 
with $n_F=0,1,2,3$.
Unfortunately, the operator of number of bosonic quanta, $n_B$, 
does not commute with $H$.
Since the model has an infinite number of bosons solving it numerically
we have to cut somewhere the Fock space off. A good choice for this cut-off 
is a maximal number of bosonic quanta in the system, $B\ge n_B$.

We are especially interested in the sector where $n_F=2$ and $j=0$.
In this sector the localized and non-localized states coexist. Moreover,
it contains the supersymmetric vacuum state. 
To find the energy spectrum we use van Baal approach 
\ci{Koller:1987fq,vanBaal:2001ac}.
For $j=0$ and $n_F=2$
the Hamiltonian can be rewritten
in terms of three bosonic variables
\begin{equation}
r^{2}=\sum_{j,a} (\hat{c}_{j}^{a})^{2}\,, \quad
u=\r^{-4}\sum_{j,a}(\hat{B}_{j}^{a})^{2}\,, \quad
v=\r^{-3}\det \hat{c} \;,
\lab{eq:ruv}
\end{equation}  
or equivalently in $(x_1,x_2,x_3)$ where
\begin{equation}
\r^{2}=\sum_{j}x_{j}^{2}\,, \quad 
u=\r^{-4}\sum_{i>j}x_{i}^{2}x_{j}^{2}\,, \quad
v=\r^{-3}\prod_{j}x_{j} \,.
\lab{eq:x123}
\end{equation}  
For example the bosonic potential has a form
\begin{equation}
H_V=u\, \r^4/2=(x_1^2 x_2^2+x_1^2 x_3^2+x_2^2 x_3^2)/2\,.
\lab{eq:Hpot}
\end{equation}
The minimum of $H_V$ is localized in six valleys along the $x_i-$axis.
The other parts of the Hamiltonian have much more complicated 
structure \ci{vanBaal:2001ac,Kotanski:2006wp}.

We construct the Fock space acting $\c_{j}^{b}$ and 
$\lambda_{\dot{\alpha}}^{a}$  on the 
empty state \ci{vanBaal:2001ac,Wosiek:2002nm} :
\begin{equation}
|n \rangle =
\sum_{\scriptsize
\begin{array}{c}
\rm{contractions}\\
\{a_1,\ldots,a_r\}
\end{array}}
\c_{k_1}^{a_1} \ldots \c_{k_m}^{a_m} 
{\bar \lambda}_{a_{m+1}}^{\dot{\alpha}} \ldots {\bar \lambda}_{a_r}^{\dot{\beta}} 
|0 \rangle\,,
\lab{eq:nstate}
\end{equation} 
where sum goes over gauge and rotation invariant combinations of the operators.
For $n_F=2$ we have two independent ways of the fermionic action:
\begin{equation}
{{\mathcal{I}}_{j}}^{a}=-2i 
\sum_{c,b,\dot{\alpha},\dot{\beta}}
\varepsilon _{c,b,a}
\bar{\lambda }_{\dot{\alpha }}^{c}
{(\bar{\sigma }^{j0})^{\dot{\alpha }}}_{\dot{\beta }}
\bar{\lambda }^{b\dot{\beta }}|0\rangle\,,
\qquad
{\mathcal{J}}^{ab}=
-\sum_{c,b,\dot{\alpha},\dot{\beta}}
\bar{\lambda }_{\dot{\alpha }}^{a}\bar{\lambda }_{\dot{\beta}}^{b}
\epsilon ^{\dot{\alpha }\dot{\beta }}|0\rangle\,,
\lab{eq:nferm}
\end{equation}  
where $\bar{\sigma }^{j0}=\half \tau_j$ and 
$\epsilon_{\alpha \beta}=\epsilon_{\dot{\alpha} \dot{\beta}}=-i \tau_2$
lowers spinor indices. 
Making contractions of (\ref{eq:nferm}) to bosons
we obtain six independent invariants
\begin{equation}
\begin{array}{ll}
|e_{1}(u,v)\rangle=\sum_{j,a} \hat{c}_{j}^{a}/\hat{r} {{\mathcal{I}}^{j}}_{a}\,, &
|e_{4}(u,v)\rangle=\sum_{a,b} \delta ^{ab} {\mathcal{J}}_{ab} \,,
\vspace{0.2cm}\\
|e_{2}(u,v)\rangle=\sum_{j,a} \hat{B}_{j}^{a}/\hat{r}^{2} {{\mathcal{I}}^{j}}_{a}\,, &
|e_{5}(u,v)\rangle=\sum_{a,b} \hat{c}_{j}^{a}\hat{c}_{j}^{b}/\hat{r}^{2} 
{\mathcal{J}}_{ab} \,,
\vspace{0.2cm}\\
|e_{3}(u,v)\rangle=\sum_{j,a} \hat{c}_{j}^{b}\hat{c}_{k}^{b}\hat{c}_{k}^{a}/\hat{r}^{3}
{{\mathcal{I}}^{j}}_{a}\,, \; &
|e_{6}(u,v)\rangle=\sum_{a,b} \hat{c}_{j}^{b}\hat{c}_{j}^{d}\hat{c}_{k}^{d}\hat{c}_{k}^{b}/\hat{r}^{4} {\mathcal{J}}_{ab} \,.
\end{array}
\lab{eq:etab}
\end{equation}  

Other basis vector can be obtained acting 
with invariant combinations of bosonic variables, \ie $(\r,u,v)$,
on these six vectors.
This gives
following basis vectors
\begin{equation}
|n \rangle 
=\sum _{m=1}^{6}h^n_{m}(\r,u,v)|e_{m}(u,v)\rangle\,,
\lab{eq:basv}
\end{equation}
where $h^n_{m}(\r,u,v)$ are arbitrary functions.
Following Refs. \ci{vanBaal:2001ac,Kotanski:2006wp} they 
are chosen as 
eigenfunctions of the harmonic oscillator.
In this basis the matrix of the kinetic Hamiltonian, $H_T$, is tridiagonal.

\section{Diagonalization of the Hamiltonian matrix}

The Hamiltonian (\ref{eq:Ham}) in the basis (\ref{eq:basv}) gives
the eigenequation
\begin{equation}
\sum_{n} H^{n' n} v_k^{n}=E_k v_k^{n'}\,,
\lab{eq:egeq}
\end{equation}
where $H^{n' n}=\langle n'| H|n\rangle$ is the Hamiltonian matrix 
\ci{vanBaal:2001ac,Kotanski:2006wp}
while $E_k$ are its eigenvalues.
The matrix for cut-off $B\le11$ is plotted in Fig. \ref{fig:HMat}.
Intensity of gray scale shows the amplitude of the matrix elements.
We can see five broad branches of the bosonic Hamiltonian,
which corresponds to change of $n_B$ by $0,\pm2,\pm4$ and
two other narrower branches of the fermionic part, $H_F$, 
where this change equals $\pm1$. 
For $B=60$ this matrix has $10416\times 10416$ elements
and its diagonalization is much more time 
consuming\footnote{The evaluation time grows exponentially with $B$.
To speed up calculation the van Baal's Mathematica 
code \ci{vanBaal:2001ac} was rewritten to C++}.

\begin{figure}[ht!]
\centerline{
\begin{picture}(90,80)
\put(5,0){\epsfysize7.6cm \epsfbox{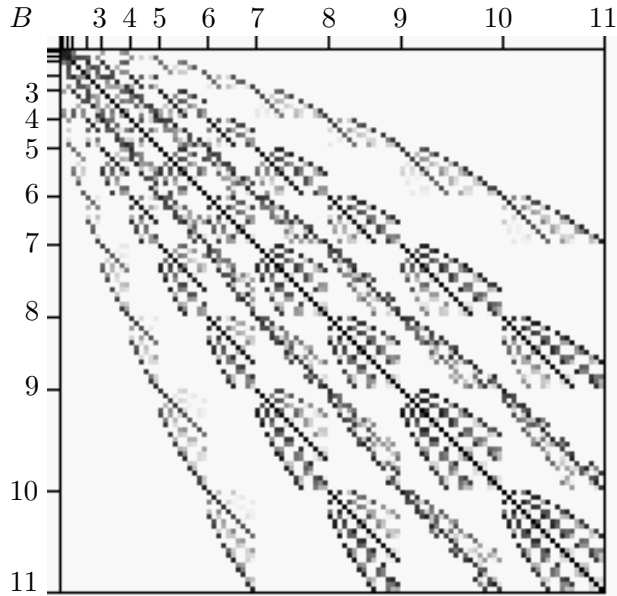}}
\put(0,77){$B$}
\put(77,77){$11$}
\put(63,77){$10$}
\put(51,77){$9$}
\put(41.5,77){$8$}
\put(32,77){$7$}
\put(25.5,77){$6$}
\put(19,77){$5$}
\put(15,77){$4$}
\put(11,77){$3$}
\put(2,67){$3$}
\put(2,63.5){$4$}
\put(2,59.5){$5$}
\put(2,53.5){$6$}
\put(2,47){$7$}
\put(2,38){$8$}
\put(2,28){$9$}
\put(0,14){$10$}
\put(0,2){$11$}
\end{picture}
}
\caption{The  $112\times 112$ Hamiltonian matrix in the base (\ref{eq:basv}) for cut-off $B\le11$.}
\lab{fig:HMat}
\end{figure}

The dependence of the energy, $E_k$, on the 
cut-off is shown in Fig. \ref{fig:Eb}.
We can see two different behaviours of the energy levels.
The first levels converge rapidly. They correspond to
localized states.
The other ones never converge but they slowly fall down 
as $E(B)\sim 1/B$.
Cut-off analyses of free particle systems gives conjecture
that the latter levels gives the continuous spectrum 
at $B\to \infty$ \ci{Trzetrzelewski:2003sz}.

\begin{figure}[ht!]
\centerline{
\begin{picture}(130,80)
\put(2,70){$E$}
\put(2,0){\epsfysize7.6cm \epsfbox{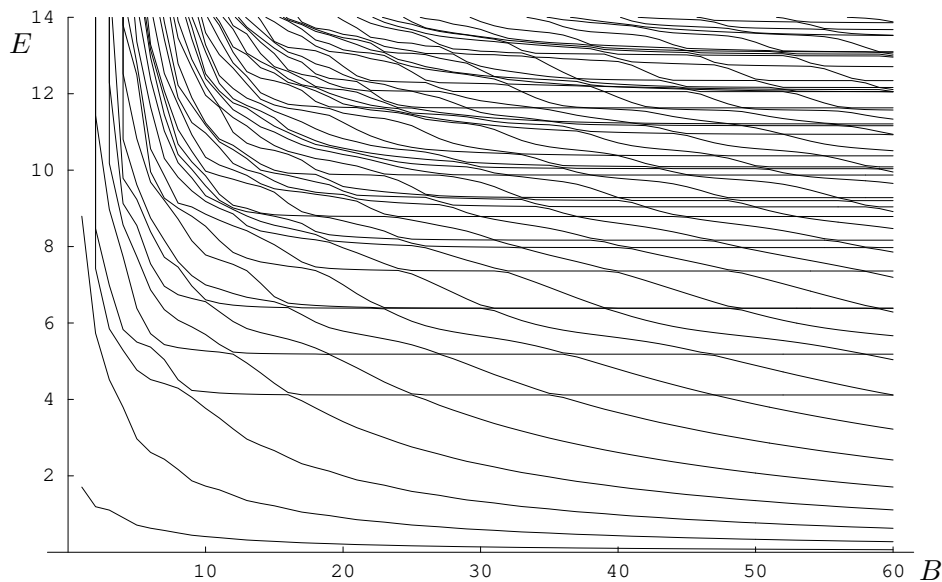}}
\put(123,0){$B$}
\end{picture}
}
\caption{Energy spectrum as a function of
cut-off $B\ge n_B$}
\lab{fig:Eb}
\end{figure}

The eigenfunctions of $H$ read 
\begin{equation}
|\Phi_k(\r,u,v)\rangle
=\sum_{n} v_k^{n} |n \rangle=\sum_{n} v_k^{n}
\sum_{m=1}^6
h^n_m(\r,u,v) |e_m \rangle\,.
\lab{eq:egfn}
\end{equation}
Examples of such wavefunctions were presented in \ci{Kotanski:2006wp}.
Indeed, the rapidly converging 
states are localized in the centre of the 
$x-$coordinate system.
On the other hand, the non-localized ones penetrate the system along 
the potential
valleys \ci{Kotanski:2006wp}. 
For these states realization of the $B \to \infty$ limit
is more complicated
and it is performed with fixed energy $E$ and changing $k$, \ie
\begin{equation}
|\Phi^{(E)}(\r,u,v)\rangle=\lim_{B \to \infty} 
|\Phi_{k(E_K=E,B)}(\r,u,v)\,\rangle.
\lab{eq:limst}
\end{equation}

Given the explicit form of the energy eigenstates
using (\ref{eq:egfn}) we can calculate averages of other operators
\begin{equation}
\langle {\cal O} \rangle_k=  \langle\Phi_k(\r,u,v)\rangle| 
{\cal O} |\Phi_k(\r,u,v)\rangle\,.
\lab{eq:avop}
\end{equation}
The only technical problem is to 
rewrite the operators in terms of $(\r,u,v)$.

\section{Virial theorem}
In order to distinguish the localized states from 
the non-localized ones
we apply the virial theorem. One can derive it from
the Heisenberg equation
\begin{equation}
\frac{d\hat{F}}{dt}
=\frac{\partial \hat{F}}{\partial t}+\frac{1}{i \hslash}[\hat{F},\hat{H}]\,.
\lab{eq:He}
\end{equation}
For a motion in a compact space the virial, 
$\vec{x}\cdot\vec{p}$, is a limited physical variable. Therefore,
its average does not change with time:
\begin{equation}
0=\frac{d}{dt} \langle  \vec{x}\cdot\vec{p}\rangle
=\frac{1}{i \hslash} \langle  [\vec{x}\cdot\vec{p},\hat{H}]\rangle\,,
\lab{eq:qvt}
\end{equation}
where the second equality follows from (\ref{eq:He}).

\begin{figure}[h!]
\centerline{
\begin{picture}(130,62)
\put(0,17){{\rotatebox{90}{\scriptsize $-2 \langle H_T \rangle 
+4 \langle  H_V \rangle + \langle H_F \rangle $}}}
\put(3,3){\epsfysize5.9cm \epsfbox{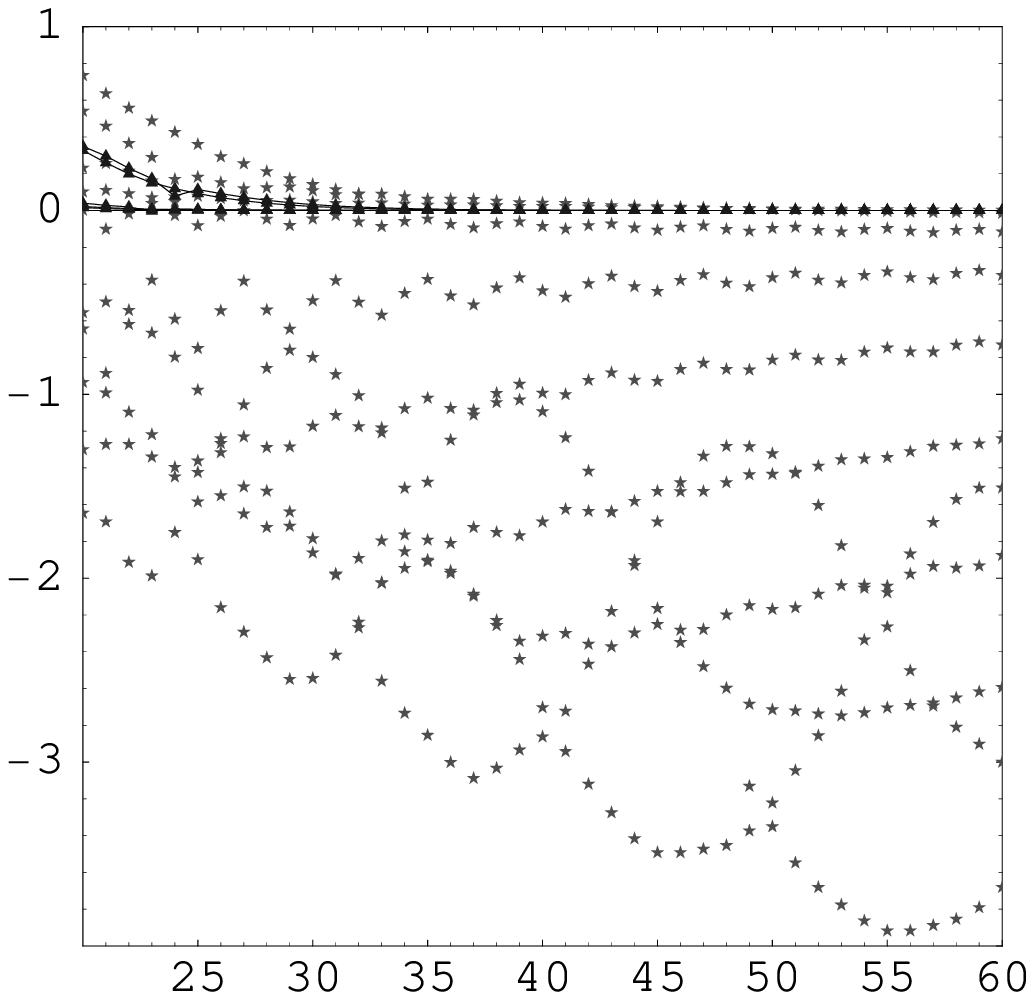}}
\put(62,17){{\rotatebox{90}{\scriptsize $-2 \langle H_T \rangle 
+4 \langle  H_V \rangle + \langle H_F \rangle $}}}
\put(65,0){\epsfysize6.5cm \epsfbox{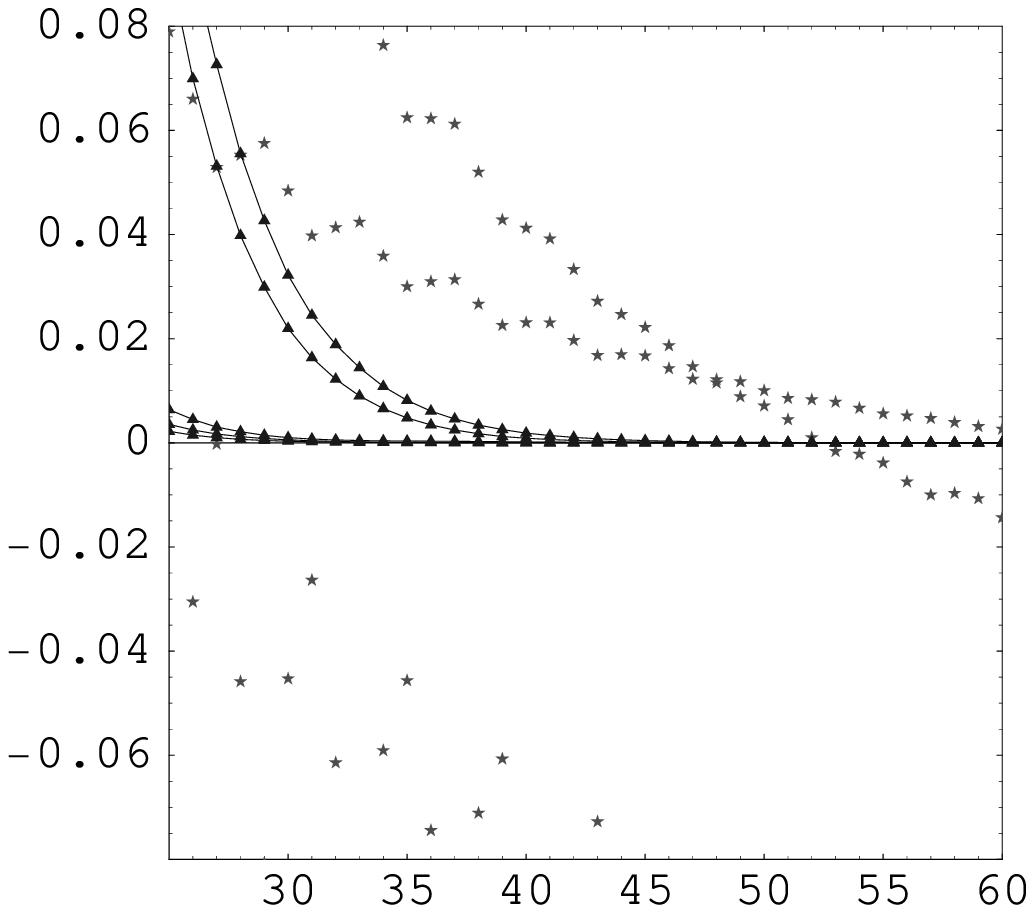}}
\put(120,0){$B$}
\put(55,0){$B$}
\end{picture}
}
\caption{The test function (\ref{eq:tfun}) 
for the bound states as a function of
cut-off $B\ge n_B$ for the the lowest fifteen energies.
Here to guide the eye joined {\it triangles} 
correspond to states from discrete spectrum
while {\it stars} are related to states with continuous spectrum}
\lab{fig:tfun}
\end{figure}

In the case of Hamiltonian $H=T+V$ where
$T$ is the kinetic energy and the potential energy scales as 
$V(\alpha \vec{x})=\alpha^n V(\vec{x})$
one obtains from (\ref{eq:qvt}):
\begin{equation}
-2 \langle T\rangle+n \langle V \rangle=0\,.
\lab{eq:toy1}
\end{equation}
In our case 
\begin{equation}
\vec{x}\cdot\vec{p} = -i 
\r \frac{\partial}{\partial \r}\,,
\lab{eq:xp}
\end{equation}
and the consecutive parts of the Hamiltonian
scale as 
\begin{equation}
H_T(\alpha \r) = \alpha^{-2} H_T(\r)
\quad
H_V(\alpha \r) = \alpha^{4} H_V(\r)
\quad
\mbox{and}
\quad
H_F(\alpha \r)=\alpha H_F(\r)\,.
\lab{eq:TVFsc}
\end{equation}
Therefore
for the bound states the test function
\begin{equation}
f \equiv -2 \langle H_T \rangle 
+4 \langle  H_V \rangle 
+ \langle H_F \rangle = 0\,.
\lab{eq:tfun}
\end{equation}

Applying (\ref{eq:avop})
the averages $\langle H_X \rangle $
are calculated over the eigenstates of 
the Hamiltonian (\ref{eq:egfn})
as
\begin{equation}
E_X =
\langle H_X \rangle_k =
\langle \Phi_k(\r,u,v)
|H_X| \Phi_k(\r,u,v)
\rangle \,.
\lab{eq:Ex}
\end{equation}

Substituting the obtained values of the averages to (\ref{eq:tfun}) 
we can check the 
relation (\ref{eq:tfun}) for the bound states. We plot the
test function (\ref{eq:tfun}) as a function of the cut-off $f(B)$
in Fig. \ref{fig:tfun}.
One can easily see that for the localized states
the relation (\ref{eq:tfun})
is fulfilled with a very good approximation
even for not so high $B$.

For the states with the continuous spectrum,
condition (\ref{eq:tfun})
is not satisfied. This is seen in Fig. \ref{fig:tfun}
where we compare the test functions 
for the bound states and the states related to the continuous spectrum
as a function of
cut-off $B\ge n_B$.
The test function (\ref{eq:tfun}) related to 
the bound states are nearly equal to zero
while the test function for the states which form the continuous spectrum
seem to grow with $B$.

\section{Summary}

In this work we discuss supersymmetric
Yang-Mills quantum mechanics (SYMQM)
\ci{Witten:1981nf,Claudson:1984th} 
in four dimensions for SU(2) gauge group.
Investigation of such models allow to understand
complicated properties of the supersymmetric theories,
\ie coexistence of localized and non-localized states,
non-triviality of the supersymmetric vacuum,
which are common for different supersymmetric theories.

We focus on the sector with the number of fermionic quanta
$n_F=2$ and the total angular momentum $j=0$.
This sector possess both discrete and continuous spectrum
\ci{deWit:1988ct},
and the supersymmetric vacuum state 
\ci{vanBaal:2001ac,Wosiek:2002nm,Kotanski:2006wp}.
In order to find the energy spectrum
we use a method proposed by van Baal in Ref. \ci{vanBaal:2001ac}
where the cut-off of our Fock space, \ie  $B$, is defined as a 
maximal number of bosonic quanta, $n_B$.
To confirm localization of states
the virial theorem is used (\ref{eq:qvt}).

The result of the Hamiltonian spectrum agree with the previous works
\ci{Wosiek:2002nm,vanBaal:2001ac,Campostrini:2004bs}.
However, here the calculations have been performed for 
the very high cut-off $B=60$. 
For this cut-off not only the spectra of bound states converge
but also the corresponding eigenstates \ci{Kotanski:2006wp}.

Our calculation shows that the quantum virial theorem 
is applicable for the systems with more complicated potential, 
\eg 
one considered in this work which consists of two
different parts.
Moreover, the virial theorem can be used to determinate localization of 
the states 
where it is not possible to calculate the Hamiltonian eigenstates
and where the averages of operators can be computed in a different way.
Thus in some cases the virial method
as the localization test of states is easier and more
applicable that direct calculation.

In the future using this method one can calculate 
matrix representations for other operators  (\ref{eq:avop}) and 
test various properties and laws for this model.
Furthermore, using similar methods one can also try to solve the models
in more dimensions and for different gauge groups
\ci{Veneziano:2005qs}.

\section*{Acknowledgements}

I thank Jacek Wosiek for 
suggesting the subject and fruitful discussions
as well as  Pierre van Baal for
making the program for calculation the energy spectrum available for me. 
I also acknowledge
discussions with Maciej Trzetrzelewski.
This work was supported by the 
grant of the Polish Ministry of Science and Education
P03B 024 27 (2004-2007).


\begin{thebibliography}{10}

\bibitem{Witten:1981nf}
E. Witten,
\newblock {\em Nucl. Phys.}, B188:513, 1981.

\bibitem{Claudson:1984th}
M. Claudson and M. B. Halpern,
\newblock {\em Nucl. Phys.}, B250:689, 1985.

\bibitem{Banks:1996vh}
T. Banks, W. Fischler, S. H. Shenker, and L. Susskind,
\newblock {\em Phys. Rev.}, D55:5112--5128, 1997;
\newblock
D. Bigatti and Leonard Susskind,
\newblock 
\newblock {\tt hep-th/9712072}, 1997;
\newblock
W. Taylor,
\newblock {\em Rev. Mod. Phys.}, 73:419--462, 2001.

\bibitem{Wosiek:2002nm}
J. Wosiek,
\newblock {\em Nucl. Phys.}, B644:85--112, 2002;
J. Kotanski and J. Wosiek,
\newblock {\em Nucl. Phys. Proc. Suppl.}, 119:932--934, 2003.

\bibitem{deWit:1988ct}
B. de Wit, M. Luscher and H. Nicolai,
\newblock {\em Nucl. Phys.}, B320:135, 1989;
\newblock
H. Nicolai and R. Helling,
\newblock {\tt hep-th/9809103}, 1998

\bibitem{Halpern:1997fv}
M. B. Halpern and C. Schwartz,
\newblock {\em Int. J. Mod. Phys.}, A13:4367--4408, 1998.

\bibitem{Polchinski:1998rq}
J. Polchinski,
\newblock String theory,
\newblock Cambridge, UK: Univ. Pr. 1998.

\bibitem{vanBaal:2001ac}
P. van Baal,
\newblock {'The Witten index beyond the adiabatic approximation'},
\newblock the Michael Marinov Memorial Volume, 'Multiple Facets of
  Quantization and Supersymmetry', 
  World Scientific.,2001; hep-th/0112072.

\bibitem{Luscher:1982ma}
M. Luscher,
\newblock {\em Nucl. Phys.}, B219:233--261, 1983;
M. Luscher and G. Munster,
\newblock {\em Nucl. Phys.}, B232:445, 1984.

\bibitem{vanBaal:1988qm}
P. van Baal,
\newblock {\em Acta Phys. Polon.}, B20:295--312, 1989.

\bibitem{Koller:1987fq}
Jeffrey Koller and P. van Baal,
\newblock {\em Nucl. Phys.}, B302:1, 1988.

\bibitem{Kotanski:2006wp}
J. Kotanski,
\newblock {\tt hep-th/0607012}, 2006.

\bibitem{Trzetrzelewski:2003sz}
M. Trzetrzelewski and J. Wosiek,
\newblock {\em Acta Phys. Polon.}, B35:1615--1624, 1989.


\bibitem{Campostrini:2004bs}
M. Campostrini and J. Wosiek,
\newblock {\em Nucl. Phys.}, B703:454--498, 2004.

\bibitem{Veneziano:2005qs}
G. Veneziano and J. Wosiek,
\newblock {\em JHEP}, 01:156, 2006;
\newblock
G. Veneziano and J. Wosiek,
\newblock {\tt hep-th/0603045}, 2006;
\newblock
E. Onofri, G. Veneziano, and J. Wosiek,
\newblock {\tt math-ph/0603082}, 2006.

\end{thebibliography}

\end{document}